\documentclass{moriond}

\def\eslt{\not\!\!{E_T}}
\def\to{\rightarrow}

\def\bi{\begin{itemize}}
\def\ei{\end{itemize}}

\def\tchi{\tilde\chi}

\def\sps1ap{SPS1a$^\prime$}
\def\c1p{C1$^\prime$}

\def\tst{\tilde t}

\def\tg{\tilde g}

\def\tell{\tilde\ell}
\def\tq{\tilde q}

\def\alt{<\sim}
\def\agt{>\sim}
\def\be{\begin{equation}}  
\def\ee{\end{equation}}  
\def\bea{\begin{eqnarray}}  
\def\eea{\end{eqnarray}}  
\def\beas{\begin{eqnarray*}}  
\def\eeas{\end{eqnarray*}}

\def\be{\begin{equation}}
\def\ee{\end{equation}}
\def\bea{\begin{eqnarray}}
\def\eea{\end{eqnarray}}

\newcommand{\Photo}{\includegraphics[height=35mm,angle=-90]{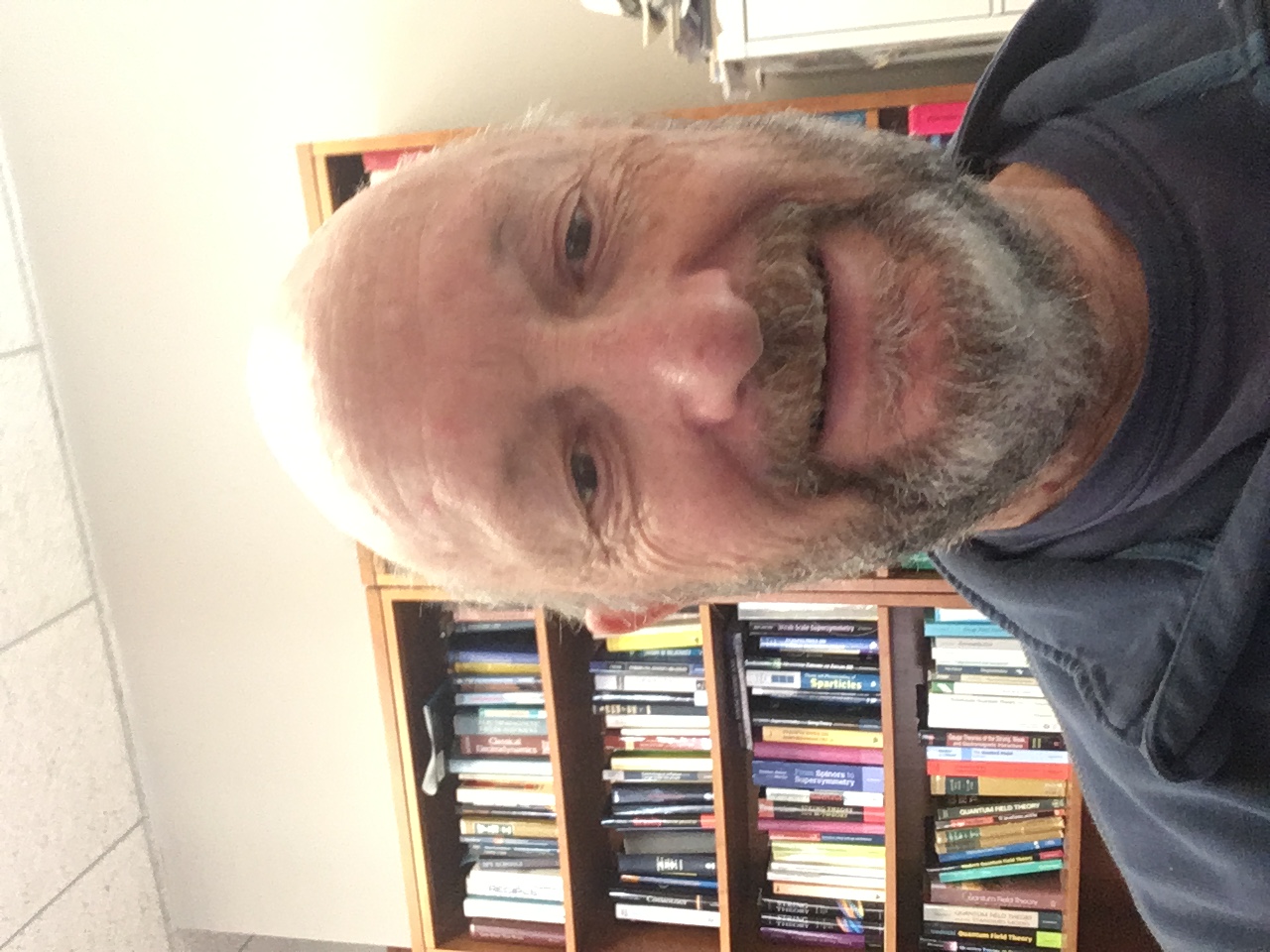}}

\begin{document}
\vspace*{4cm}
\title{Beyond the Standard Model: An overview}

\author{Howard Baer}

\address{Department of Physics and Astronomy,\\
  University of Oklahoma, Norman OK 73071 USA}

\maketitle\abstracts{At present, the Standard Model (SM) agrees with almost all
  collider data. Yet, three finetuning issues-- the Higgs mass problem,
  the strong CP problem and the cosmological constant problem--
  all call for new physics. The most plausible solutions at present are
  weak scale SUSY, the PQWW axion and the string landscape.
  A re-evaluation of EW finetuning in SUSY allows for a higgsino-like LSP and
  naturalness upper bounds well beyond LHC limits. Rather general
  arguments from string theory allow for statistical predictions that
  $m_h\sim 125$ GeV with sparticles beyond present LHC limits.
  The most lucrative LHC search channel may be for light higgsino pair
  production.
  Dark matter turns out to be a SUSY DFSZ axion along with a diminished
  abundance of higgsino-like WIMPs.
}

I would like to present an overview of physics Beyond the Standard Model
(BSM). But with only 20 minutes, I will stick to what I regard as the most
plausible avenues for discovery based on our present knowledge of
particle theory. I'll take Weinberg's advice, and focus on the so-called
``naturalness problems'', and Einstein's advice, and seek parsimonius
solutions to these. We have seen that a wide variety of collider and other
data are in spectacular agreement with SM predictions, and many previous
anomalies, including $(g-2)_\mu$, either went away or seem to be going away.
And yet, it is clear that the SM is not the final word in particle theory.

\section{Naturalness and the gauge hierarchy, strong CP and CC problems}

The SM is beset by three naturalness problems: 1. the gauge hierarchy,
or Higgs mass problem, 2. the strong CP problem and 3. the cosmological constant (CC) problem.

The most plausible solutions to these are: 1. weak scale supersymmetry\cite{Baer:2006rs} (SUSY), 2. the PQWW\cite{DiLuzio:2020wdo} axion and 3. Weinberg's
anthropic solution\cite{Weinberg:1987dv} to the CC problem which emerges
from the vast landscape of string vacua\cite{Douglas:2019kus}.

\section{Electroweak naturalness and SUSY}
\label{sec:nat}

SUSY provides a 't Hooft technically natural solution to the so-called
Big Hierarchy Problem (BHP), but our concern these days is with the
Little Hierarchy Problem (LHP): why is the scale of soft SUSY breaking terms
$m_{soft}$ well above the measured value of the weak scale
$m_{weak}\sim m_{W,Z,h}\sim 100$ GeV? 
Current LHC constraints require $m_{\tg}\agt 2.2$ TeV and $m_{\tst_1}\agt 1.1$ TeV.
This may be compared with early upper limits from naturalness\cite{Dimopoulos:1995mi} for no
finetuning below the 3\% level that $m_{\tg}$ and $m_{\tst_1}\alt 400$ GeV.
Furthermore, the promised WIMPs have yet to show up at WIMP direct detection
experiments\cite{LZ:2022lsv}.
From Ref. \cite{Arkani-Hamed:2015vfh}: ``settling the ultimate fate of
naturalness is perhaps the most profound theoretical question of our time''.
This involves some scrutiny of the commonly used naturalness measures.

The LHP is instead as a matter of {\it practical naturalness:}\cite{Baer:2015rja} an observable
${\cal O}$ is practically natural if all independent contributions to
${\cal O}$ are comparable to or less than ${\cal O}$. Now in the MSSM,
minimization of the scalar potential leads to
\be
m_Z^2/2=\frac{m_{H_d}^2+\Sigma_d^d-(m_{H_u}^2+\Sigma_u^u)\tan^2\beta}{\tan^2\beta -1}-\mu^2\simeq -m_{H_u}^2-\Sigma_u^u(\tst_{1,2})-\mu^2
\label{eq:mzs}
\ee
where $m_{H_u}^2$ and $m_{H_d}^2$ are soft breaking Higgs masses, $\mu$ is the
SUSY conserving $\mu$ parameter and the $\Sigma_{u,d}^{u,d}$ terms contain over 40 one-loop and some two-loop corrections\cite{Baer:2012cf}
(as programmed in Isajet/DEW4SLHA\cite{Paige:2003mg}).
The measure $\Delta_{EW}$\cite{Baer:2012up} compares
the largest term on the rigt-handside against $m_Z^2/2$:
if any term is far greater than $m_Z^2/2$, then some other would have to be
opposite-sign finetuned to gain $m_Z=91.2$ GeV, its measured value.
Such tunings are highly implausible.
Eq. \ref{eq:mzs} is where the EW finetuning is actually implemented
in spectra codes such as Isajet, SoftSUSY etc.
As such it is the most conservative of naturalness measures,
and is equally applicable to high scale SUSY models or the pMSSM. 

An older measure,
$\Delta_{p_i}\equiv max_i|(p_i/m_Z^2)\partial m_Z^2 /\partial p_i|$,
determines how the weak scale changes against variation in model parameters
$p_i$ which are usually taken as different high scale soft terms.
Here, the debacle is that the $p_i$ parametrize our ignorance of our model of
SUSY breaking. But nature isn't ignorant. In particular SUSY breaking models,
the soft parameters (in our universe) are all correlated and not
independent\cite{Baer:2013gva,Mustafayev:2014lqa}.
This leads to overestimates of finetuning by factors of 10-1000\cite{Baer:2023cvi}.

A second measure $\Delta_{HS}\equiv \delta m_h^2/m_h^2$ with
$\delta m_h^2\sim \delta m_{H_u}^2\sim -3\lambda_t^2/(8\pi^2)(m_{Q_3}^2+m_{U_3}^2+A_t^2)\log (\Lambda/m_{SUSY} )$ demands three third generation squarks below
$\sim 500$ GeV for $\Lambda\sim m_{GUT}$. This measure arises from solving the
soft term RGE for $m_{H_u}^2$ but under the simplification of setting gauge terms, the $S$ term and $m_{H_u}^2$ itself to zero.
The latter simplification betrays that $m_{H_u}^2(\Lambda )$ and
$\delta m_{H_u}^2$ are not really independent, and shouldn't be tuned one
against the other. This also leads to overestimates of finetuning by
factors $10-1000$\cite{Baer:2023cvi}.

Using $\Delta_{EW}\alt 30$, then the implications are that $\mu\alt 350$ GeV
so we get a higgsino-like LSP. Top squarks can range up to $3$ TeV and gluinos
up to $6$ TeV\cite{Baer:2018hpb}.
Using $\Delta_{EW}$, there is no naturalness crisis for
SUSY under present LHC search limits.

\section{SUSY from the string landscape}
\label{sec:land}

The emergence of the string landscape\cite{Bousso:2000xa} provides a setting for Weinberg's
anthropic solution of the CC problem. Under flux compactifications\cite{Douglas:2006es},
string theory may admit up to $10^{500}-10^{1000}$ different vacuum
configurations, each leading different $4-d$ laws of physics.
In particular, we expect the CC and the SUSY breaking scale $m_{hidden}^2$
to scan in the landscape. Weinberg's suggestion is that the CC should have
a non-zero value, but only those vacua with $\rho_{vac}\alt 10^{-123} m_P^4$
should allow for galaxy condensation, and hence large scale structure formation,
in different pocket-universes (PU) within the greater multiverse.

Likewise, the SUSY breaking scale $m_{hidden}$ is expected to scan\cite{Arkani-Hamed:2005zuc} within the
multiverse. Since there seems to be nothing in string theory favoring
any particular value of $m_{hidden}$, then Douglas suggested\cite{Douglas:2004qg} that
$F$-breaking fields be distributed randomly as complex numbers whilst
$D$-breaking fields are distributed as real numbers. The ultimate SUSY breaking scale
\be
m_{hidden}^4=\sum_i F^\dagger F+\sum_\alpha D_\alpha D_\alpha
\ee
should then be distributed in the multiverse according to the volume of
the outer shells in SUSY breaking space:
\be
f_{SUSY}\sim m_{soft}^{2n_F+n_D-1}
\ee
where $m_{soft}\sim m_{hidden}^2/m_P$,
  $n_F$ is the number of hidden sector $F$-breaking fields and $n_D$ is
the number of $D$-breaking fields contributing to the overall SUSY breaking scale. Thus, even for the textbook case of SUSY breaking via a single $F$-term field, one expects a linear draw in the multiverse to large soft terms.
Regarding the soft terms, the gaugino masses, various scalar masses and trilinears $A$ are expected to scan independently due to their different functional
dependencies on hidden sector fields associated with SUSY breaking\cite{Baer:2020vad}.

The draw to large soft terms must be balanced by the derived value for the weak scale in each PU lying within the ABDS window of allowed
values\cite{Agrawal:1997gf}
\be
m_{weak}^{PU}: 50\ {\rm GeV}\sim 350\ {\rm GeV}\ \ \ ({\rm ABDS\ window})
\ee
lest complex nuclei, and hence atoms as we know them, not form.
The upper limit on the ABDS window corresponds to
$m_{weak}^{PU}\alt m_Z\sqrt{\Delta_{EW}/2}$ absent finetuning in each PU.
Combining the draw to large soft terms $f_{SUSY}$ with the ABDS anthropic
window, one can scan over models such as NUHM2-4\cite{Ellis:2002iu},
generalized mirage mediation\cite{Baer:2016hfa} and
natural generalized anomaly-mediation\cite{Baer:2018hwa} to generate stringy probability distributions for Higgs and sparticle masses (scaling the value of
$m_Z^{PU}\to 91.2$ GeV to gain the mass distribution expected in our universe).
\begin{figure}
\begin{minipage}{0.30\linewidth}
\includegraphics[width=1.5\linewidth]{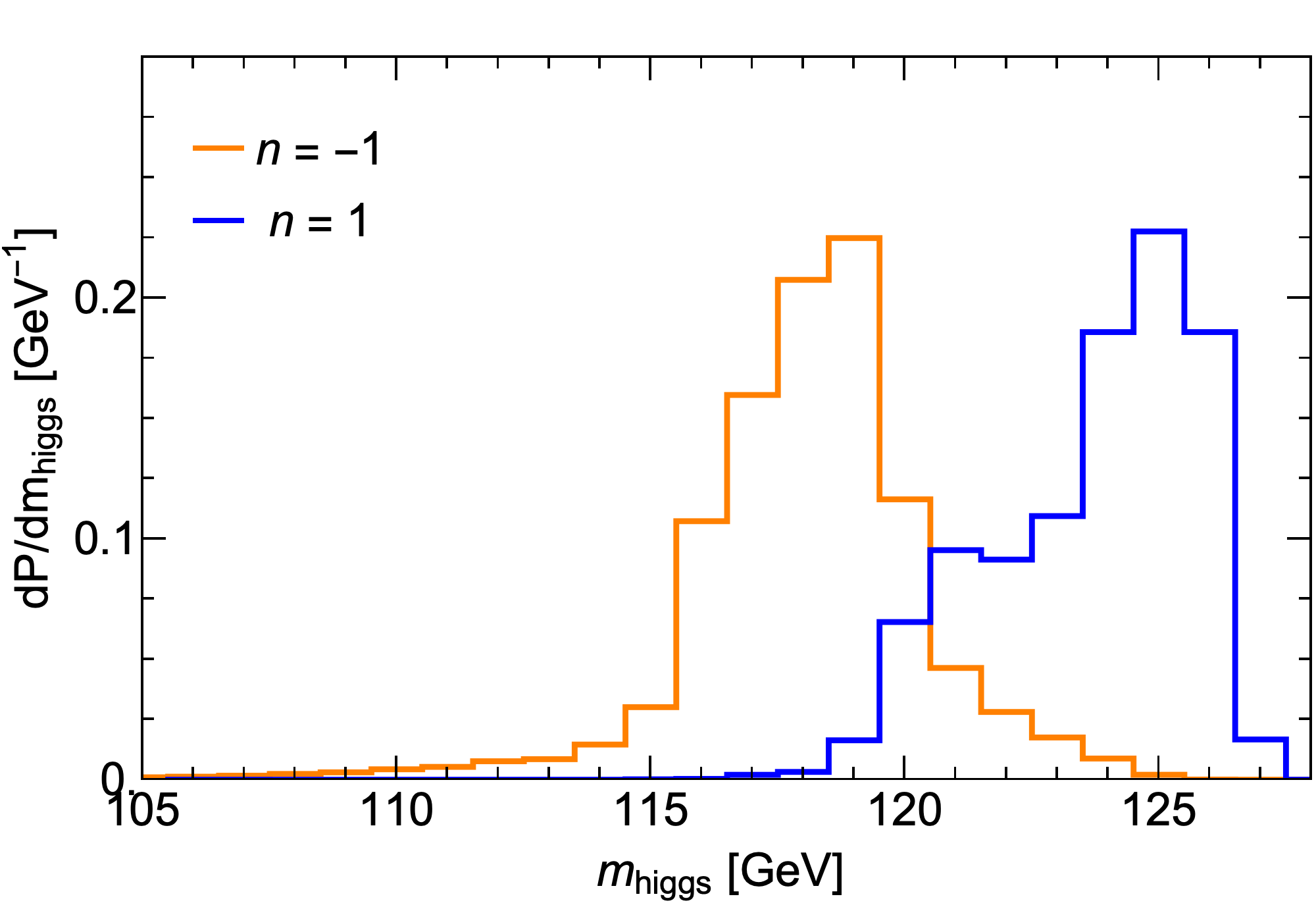}\centering
\end{minipage}
\caption[]{Statistical prediction of $m_h$ from the string landscape
  for an $n=\pm 1$ power-law draw on soft SUSY breaking terms in the
  NUHM model.
}
\label{fig:mhiggs}
\end{figure}

The Higgs mass distribution is shown in Fig. \ref{fig:mhiggs} for
$n\equiv 2n_F+n_D-1 =\pm 1$. The linear draw to large soft terms
yields a maximal $m_h$ expectation that $m_h\sim 125$ GeV while
gluinos live in the range $m_{\tg}\sim 2-6$ TeV and top squarks in the range
$m_{\tst_1}\sim 1-1.2$ TeV.
First/second generation squarks and sleptons have Yukawa suppressed
contributions to $m_{weak}$ and under RG running actually contribute via
two-loop RG terms that depend on their gauge couplings. They are preferred
in the $m_{\tq,\tell}\sim 10-40$ TeV range, yielding a mixed
decoupling/quasi-degeneracy landscape solution to the SUSY flavor and CP problems\cite{Baer:2019zfl}.

\begin{figure}
\begin{minipage}{0.30\linewidth}
\includegraphics[width=1.5\linewidth]{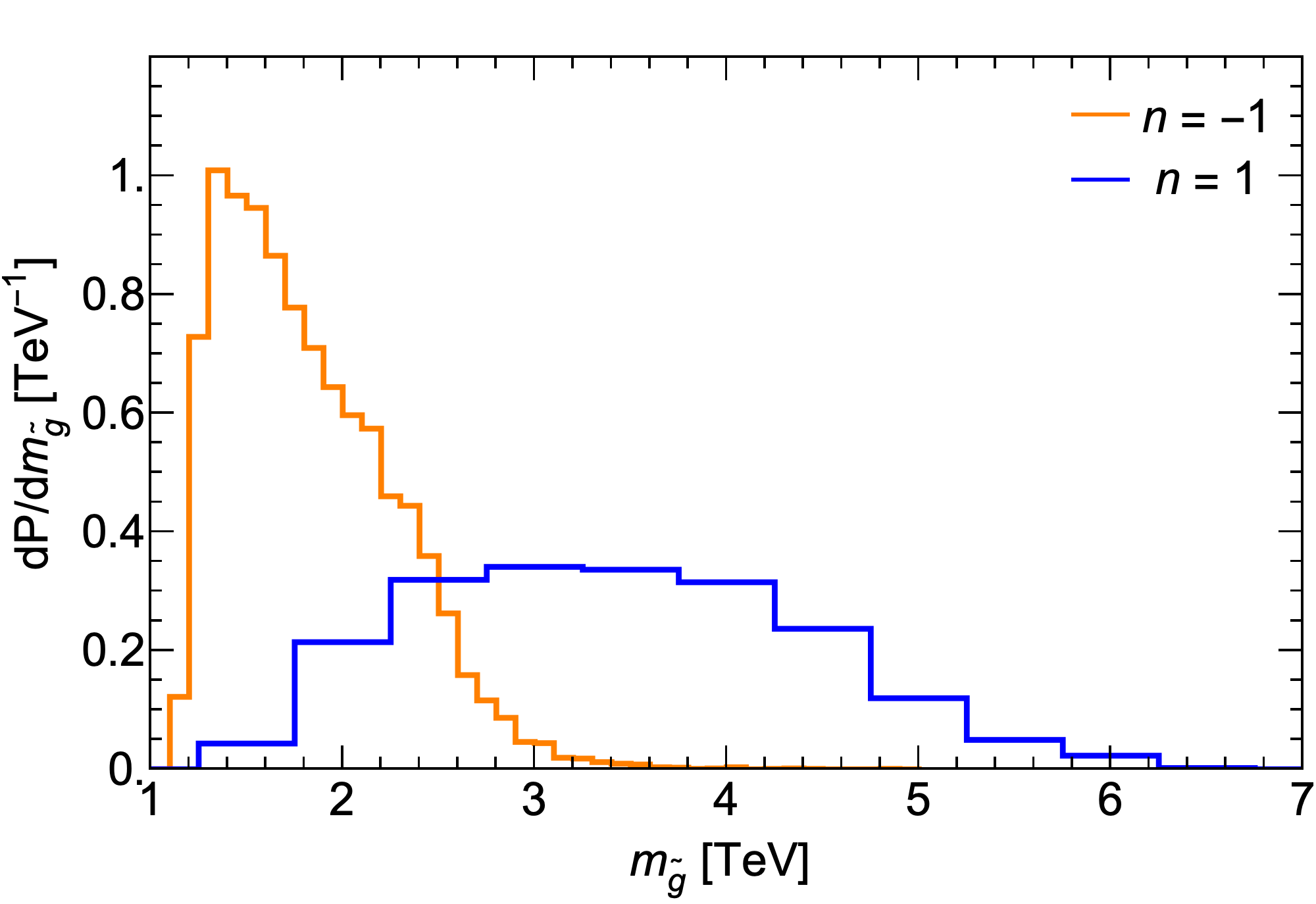}\centering
\end{minipage}
\caption[]{Statistical prediction of $m_{\tg}$ from the string landscape
  for an $n=\pm 1$ power-law draw on soft SUSY breaking terms in the
  NUHM model.}
\label{fig:mgl}
\end{figure}

From this stringy naturalness perspective\cite{Baer:2019cae}, LHC experiments are only {\it beginning} to probe the expected
range of sparticle masses. The prediction from the string landscape
is that $m_h\simeq 125$ GeV with sparticles somewhat or well-beyond
present LHC search limits: just what LHC is seeing! Under stringy
naturalness, a 3 TeV gluino is more natural than the 300 GeV gluino:
see Fig. \ref{fig:mgl}.

\section{LHC searches for landscape SUSY}
\label{sec:lhc}

\subsection{Higgsino pair production}

Since $\mu$ enters directly into Eq. \ref{eq:mzs}, and is SUSY conserving,
it feeds mass to $W$, $Z$ and $h$ and also the four light higgsinos
$\tchi_1^\pm$, $\tchi_{1,2}^0$. The higgsino mass splittings are typically
5-15 GeV in natSUSY models.
Thus, heavier higgsinos decay to lighter higgsinos
plus fermion-antifermion with most of the rest energy of the mother particle
going to the daughter LSP $m_{\tchi_1^0}$.
The visible decay products are very soft making higgsino pair production
very challenging at LHC\cite{Baer:2011ec}.
A way forward is to look for boosted $\ell\bar{\ell}$ from
$\tchi_2^0\to\tchi_1^0 \ell\bar{\ell}$ where the higgsino pair recoils
from a hard initial state jet radiation\cite{Han:2014kaa,Baer:2014kya}.
The $m(\ell\bar{\ell})$ is kinematically
bounded by $m_{\tchi_2^0}-m_{\tchi_1^0}$ so the dilepton mass excess should
be confined to the lower invariant mass bins. This signal can be augmented
by $pp\to \tchi_1^\pm\tchi_2^0$ where $\tchi_1^\pm\to\tchi_1^0f\bar{f}^\prime$
but the charged higgsino decay products may be even softer than those from
from $\tchi_2^0$ due to the model dependence of the loop corrections to the
higgsino mass states: thus, caution should be used in combining these signals.
There is at present some excess in the soft dilepton $+jets +\eslt$ channel
for both ATLAS\cite{ATLAS:2019lng} and CMS\cite{CMS:2021edw} and one must see if the excess is enhanced under
further data taking in Run 3 and beyond.

\subsection{Wino pair production}

In natSUSY, one may also look for wino pair production followed by two body
decay to $W$, $Z$ or $h$ plus higgsino. A new analysis of $VV+\eslt$
production followed by boosted hadronic $V=W,Z$ decay excludes some higher
$m(wino)$ mass range\cite{ATLAS:2021yqv} but the reaction $pp\to\tchi_2^\pm\tchi_{3,4}^0$
production can lead to a robust same-sign diboson signature\cite{Baer:2013yha}
$W^\pm W^\pm +\eslt$
which is still awaiting a dedicated experimental search.

\subsection{Top-squark pair production}

Top-squark pair production $pp\to\tst_1\tst_1^*$ is also lucrative in natSUSY.
A recent analysis\cite{Baer:2023uwo} claims HL-LHC has a reach out to $m_{\tst_1}\sim 1.7$ TeV
which would cover about half the expected range from the string landscape.

\section{SUSY dark matter: an axion, higgsinolike WIMP admixture}
\label{sec:dm}

It is well-known\cite{Baer:2013vpa} that higgsino-like LSPs are thermally underproduced from
the measured dark matter relic density by typically a factor $\sim 10$.
However, in respecting naturalness, one also likely needs the PQ solution
to the strong CP problem, namely the QCD axion $a$. In this case, one has
multi-component dark matter which is an $a-\tchi_1^0$ admixture\cite{Baer:2011hx}.
The DFSZ axion, like SUSY, requires two Higgs doublets and so meshes well
with SUSY\cite{Bae:2013hma}.
In fact, axions solve several problems in SUSY while SUSY solves
several problems with axion physics (introducing a new scale
$f_a\sim 10^{11}$ GeV requires more fervently stabilization of the
Higgs sector via SUSY).

The required global $U(1)_{PQ}$ is incompatible with quantum gravity and
hence is expected to be an approximate, accidental symmmetry arising from
something which is gravity compatible. A lucrative possibility is a discrete
$R$-symmetry ${\bf Z}_n^R$ which can arise from string compactifications.
The anomaly-free ${\bf Z}_n^R$ symmetries compatible with GUT
representations have been catalogued by Lee {\it et al.}\cite{Lee:2011dya}
and the largest of these,
${\bf Z}_{24}^R$, is found to forbid non-renormalizable operators up to
$(1/m_P)^7$ in the superpotential for two extra field models $X$ and $Y$.
This solves the axion quality problem while also generating the required
global PQ, generating $R$-parity conservation, generating a SUSY $\mu$
term via the Kim-Nilles mechanism and forbidding dangerous proton decay
operators\cite{Baer:2018avn}.
The PQ scale $f_a$ is related to the SUSY breaking scale
and hence in the cosmological sweet spot: $f_a\sim m_{hidden}\sim 10^{11}$ GeV.

To compute the DM relic density, one now needs eight coupled Boltzmann equations which account for axino, gravitino, neutralino and thermal and
non-thermal axion and saxion production along with radiation\cite{Bae:2014rfa}.
Typically one gets mainly SUSY DFSZ axion dark matter with a smattering
of higgsino-like WIMPs\cite{Bae:2013bva}.
The $a\gamma\gamma$ coupling is reduced within the MSSM
by the presence of higgsinos circulating in the triangle diagram\cite{Bae:2017hlp}.
The WIMP direct detection rates are diminished by a factor $\xi\equiv \Omega_{\tchi_1^0}/0.12$ while indirect detection rates from WIMP-WIMP annihilation are suppressed by $\xi^2$ so this scenario is still viable\cite{Baer:2016ucr} even in light of
multi-ton Xenon detectors such as Xenon-n-ton and LZ\cite{Baer:2016ucr}.

Recently, the added effect of light stringy moduli has been included, requiring
nine coupled Boltzmass equations\cite{Bae:2022okh}.
In natSUSY scenarios with light higgsinos,
the lightest modulus mass $m_\phi\agt 5$ PeV in order that the modulus decays
before neutralino freeze-out, to avoid overclosing the universe\cite{Baer:2023bbn}.

\section{Conclusions}
\label{sec:conclude}

1. Time to set aside old/flawed measures of naturalness,
2. there is plenty of natural parameter space under the model-independent
measure $\Delta_{EW}$,
3. $\mu\sim 100-350$ GeV: light higgsinos!
4. other sparticle contributions to $m_{weak}$ are loop suppressed-
masses can be TeV$\to$ multi-TeV,
4. stringy naturalness (SN): what the string landscape prefers,
5. landscape draw to large soft terms provided
$m_{weak}^{PU}\sim (2-5) m_{weak}^{OU}$ GeV,
6. SN predicts LHC sees $m_h\sim 125$ GeV but as yet no sign of sparticles,
7. under stringy naturalness, a 3 TeV gluino is more natural than
300 GeV gluino,
8. string landscape $\Rightarrow$ non-universal 1st/2nd gen.
scalars at 20-40 TeV: natural but gives quasi-degeneracy/decoupling
sol’n to SUSY flavor/CP problems,
9. most promising at LHC: light higgsinos via soft   dilepton$+jet+\eslt$
channel,
10. ILC: for $\sqrt{s}>2m(higgsino)$, expect a higgsino factory and
11. dark matter: a mix of axions plus higgsino-like WIMPs
(typically mainly axions).

\section*{Acknowledgments}

I thank Farvah Mahmoudi for her kind invitation to speak at Moriond QCD 2024.
I thank my collaboraters for their contributions to this overview.


\section*{References}

\begin{thebibliography}{99}

\bibitem{Baer:2006rs} H.~Baer and X.~Tata,
``Weak scale supersymmetry: From superfields to scattering events,''
Cambridge University Press, 2006.
  
\bibitem{DiLuzio:2020wdo}
L.~Di Luzio, M.~Giannotti, E.~Nardi and L.~Visinelli,
Phys. Rept. \textbf{870} (2020), 1-117.

\bibitem{Weinberg:1987dv}
S.~Weinberg,
Phys. Rev. Lett. \textbf{59} (1987), 2607
doi:10.1103/PhysRevLett.59.2607

\bibitem{Douglas:2019kus}
M.~R.~Douglas,
Universe \textbf{5} (2019) no.7, 176
doi:10.3390/universe5070176

\bibitem{Dimopoulos:1995mi}
S.~Dimopoulos and G.~F.~Giudice,
Phys. Lett. B \textbf{357} (1995), 573-578.

\bibitem{LZ:2022lsv}
J.~Aalbers \textit{et al.} [LZ],
Phys. Rev. Lett. \textbf{131} (2023) no.4, 041002.

\bibitem{Arkani-Hamed:2015vfh}
N.~Arkani-Hamed, T.~Han, M.~Mangano and L.~T.~Wang,
Phys. Rept. \textbf{652} (2016), 1-49.

\bibitem{Baer:2015rja}
H.~Baer, V.~Barger and M.~Savoy,
Phys. Rev. D \textbf{93} (2016) no.3, 035016.

\bibitem{Baer:2012cf}
H.~Baer, V.~Barger, P.~Huang, D.~Mickelson, A.~Mustafayev and X.~Tata,
Phys. Rev. D \textbf{87} (2013) no.11, 115028.

\bibitem{Paige:2003mg}
F.~E.~Paige, S.~D.~Protopopescu, H.~Baer and X.~Tata,
[arXiv:hep-ph/0312045 [hep-ph]].

\bibitem{Baer:2012up}
H.~Baer, V.~Barger, P.~Huang, A.~Mustafayev and X.~Tata,
Phys. Rev. Lett. \textbf{109} (2012), 161802.

\bibitem{Baer:2013gva}
H.~Baer, V.~Barger and D.~Mickelson,
Phys. Rev. D \textbf{88} (2013) no.9, 095013.

\bibitem{Mustafayev:2014lqa}
A.~Mustafayev and X.~Tata,
Indian J. Phys. \textbf{88} (2014), 991-1004.

\bibitem{Baer:2023cvi}
H.~Baer, V.~Barger, D.~Martinez and S.~Salam,
Phys. Rev. D \textbf{108} (2023) no.3, 035050.

\bibitem{Baer:2018hpb}
H.~Baer, V.~Barger, J.~S.~Gainer, D.~Sengupta, H.~Serce and X.~Tata,
Phys. Rev. D \textbf{98} (2018) no.7, 075010.

\bibitem{Bousso:2000xa}
R.~Bousso and J.~Polchinski,
JHEP \textbf{06} (2000), 006.

\bibitem{Douglas:2006es}
M.~R.~Douglas and S.~Kachru,
Rev. Mod. Phys. \textbf{79} (2007), 733-796.

\bibitem{Arkani-Hamed:2005zuc}
N.~Arkani-Hamed, S.~Dimopoulos and S.~Kachru,
[arXiv:hep-th/0501082 [hep-th]].

\bibitem{Douglas:2004qg}
M.~R.~Douglas,
[arXiv:hep-th/0405279 [hep-th]].

\bibitem{Baer:2020vad}
H.~Baer, V.~Barger, S.~Salam and D.~Sengupta,
Phys. Rev. D \textbf{102} (2020) no.7, 075012.

\bibitem{Agrawal:1997gf}
V.~Agrawal, S.~M.~Barr, J.~F.~Donoghue and D.~Seckel,
Phys. Rev. D \textbf{57} (1998), 5480-5492.

\bibitem{Ellis:2002iu}
J.~R.~Ellis, T.~Falk, K.~A.~Olive and Y.~Santoso,
Nucl. Phys. B \textbf{652} (2003), 259-347.

\bibitem{Baer:2016hfa}
H.~Baer, V.~Barger, H.~Serce and X.~Tata,
Phys. Rev. D \textbf{94} (2016) no.11, 115017.

\bibitem{Baer:2018hwa}
H.~Baer, V.~Barger and D.~Sengupta,
Phys. Rev. D \textbf{98} (2018) no.1, 015039.

\bibitem{Baer:2019zfl}
H.~Baer, V.~Barger and D.~Sengupta,
Phys. Rev. Res. \textbf{1} (2019) no.3, 033179.

\bibitem{Baer:2019cae}
H.~Baer, V.~Barger and S.~Salam,
Phys. Rev. Research. \textbf{1} (2019), 023001.

\bibitem{Baer:2011ec}
H.~Baer, V.~Barger and P.~Huang,
JHEP \textbf{11} (2011), 031.

\bibitem{Han:2014kaa}
Z.~Han, G.~D.~Kribs, A.~Martin and A.~Menon,
Phys. Rev. D \textbf{89} (2014) no.7, 075007.

\bibitem{Baer:2014kya}
H.~Baer, A.~Mustafayev and X.~Tata,
Phys. Rev. D \textbf{90} (2014) no.11, 115007.

\bibitem{ATLAS:2019lng}
G.~Aad \textit{et al.} [ATLAS],
Phys. Rev. D \textbf{101} (2020) no.5, 052005.

\bibitem{CMS:2021edw}
A.~Tumasyan \textit{et al.} [CMS],
JHEP \textbf{04} (2022), 091.

\bibitem{ATLAS:2021yqv}
G.~Aad \textit{et al.} [ATLAS],
Phys. Rev. D \textbf{104} (2021) no.11, 112010.

\bibitem{Baer:2013yha}
H.~Baer, V.~Barger, P.~Huang, D.~Mickelson, A.~Mustafayev, W.~Sreethawong and X.~Tata,
Phys. Rev. Lett. \textbf{110} (2013) no.15, 151801.

\bibitem{Baer:2023uwo}
H.~Baer, V.~Barger, J.~Dutta, D.~Sengupta and K.~Zhang,
Phys. Rev. D \textbf{108} (2023) no.7, 075027.

\bibitem{Baer:2013vpa}
H.~Baer, V.~Barger and D.~Mickelson,
Phys. Lett. B \textbf{726} (2013), 330-336.

\bibitem{Baer:2011hx}
H.~Baer, A.~Lessa, S.~Rajagopalan and W.~Sreethawong,
JCAP \textbf{06} (2011), 031.

\bibitem{Bae:2013hma}
K.~J.~Bae, H.~Baer and E.~J.~Chun,
JCAP \textbf{12} (2013), 028.

\bibitem{Lee:2011dya}
H.~M.~Lee, S.~Raby, M.~Ratz, G.~G.~Ross, R.~Schieren, K.~Schmidt-Hoberg and P.~K.~S.~Vaudrevange,
Nucl. Phys. B \textbf{850} (2011), 1-30.

\bibitem{Baer:2018avn}
H.~Baer, V.~Barger and D.~Sengupta,
Phys. Lett. B \textbf{790} (2019), 58-63.

\bibitem{Bae:2013bva}
K.~J.~Bae, H.~Baer and E.~J.~Chun,
Phys. Rev. D \textbf{89} (2014) no.3, 031701.

\bibitem{Bae:2014rfa}
K.~J.~Bae, H.~Baer, A.~Lessa and H.~Serce,
JCAP \textbf{10} (2014), 082.

\bibitem{Bae:2017hlp}
K.~J.~Bae, H.~Baer and H.~Serce,
JCAP \textbf{06} (2017), 024.

\bibitem{Baer:2016ucr}
H.~Baer, V.~Barger and H.~Serce,
Phys. Rev. D \textbf{94} (2016) no.11, 115019.

\bibitem{Bae:2022okh}
K.~J.~Bae, H.~Baer, V.~Barger and R.~W.~Deal,
JHEP \textbf{02} (2022), 138.

\bibitem{Baer:2023bbn}
H.~Baer, V.~Barger and R.~Wiley Deal,
JHEP \textbf{06} (2023), 083.



\end{thebibliography}
\end{document}